%% file: sample-acmsmall.tex
\documentclass[acmsmall]{acmart}

\setcopyright{none} 
\usepackage[inline]{enumitem}%
\usepackage{graphicx}%
\usepackage{multirow}%
\usepackage{mathrsfs}%
\usepackage{xcolor}%
\usepackage{multirow}% 
\usepackage[normalem]{ulem}% 
\useunder{\uline}{\ul}{}% 
\usepackage[title]{appendix}%
\usepackage{xcolor}%
\usepackage{textcomp}%
\usepackage{manyfoot}%
\usepackage{booktabs}%
\usepackage{algorithm}%
\usepackage{algorithmicx}%
\usepackage{algpseudocode}%
\usepackage{listings}%
\usepackage{fancyvrb}
\usepackage{subcaption}

\acmDOI{}

\begin{document}

\title{Leveraging Large Language Models for Generating Research Topic Ontologies: A Multi-Disciplinary Study}
\renewcommand{\shorttitle}{LLMs for Generating Research Topic Ontologies}

\author{Tanay Aggarwal}
\email{tanay.aggarwal@open.ac.uk}
\orcid{0009-0009-9477-7112}
\affiliation{%
  \institution{Knowledge Media Institute, The Open University}
  \city{Milton Keynes}
  \country{UK}
}
\authornote{Corresponding author(s).}

\author{Angelo Salatino}
\email{angelo.salatino@open.ac.uk}
\orcid{0000-0002-4763-3943}
\affiliation{%
  \institution{Knowledge Media Institute, The Open University}
  \city{Milton Keynes}
  \country{UK}
}

\author{Francesco Osborne}
\orcid{0000-0001-6557-3131}
\email{francesco.osborne@open.ac.uk}
\affiliation{%
  \institution{Knowledge Media Institute, The Open University}
  \city{Milton Keynes}
  \country{UK}}
\affiliation{
  \institution{Department of Business and Law, University of Milano Bicocca}
  \city{Milan}
  \country{IT}}

\author{Enrico Motta}
\email{enrico.motta@open.ac.uk}
\orcid{0000-0003-0015-1952}
\affiliation{%
  \institution{Knowledge Media Institute, The Open University}
  \city{Milton Keynes}
  \country{UK}
}

\renewcommand{\shortauthors}{Aggarwal et al.}
\renewcommand\footnotetextcopyrightpermission[1]{}
\fancyfoot{}
\sloppy

\begin{abstract}
Ontologies and taxonomies of research fields are critical for managing and organising scientific knowledge, as they facilitate efficient classification, dissemination and retrieval of information. 
However, the creation and maintenance of such ontologies are expensive and time-consuming tasks, usually requiring the coordinated effort of multiple domain experts. Consequently, ontologies in this space often exhibit uneven coverage across different disciplines, limited inter-discipline connectivity, and infrequent updating cycles.
In this study, we investigate the capability of several large language models to identify semantic relationships among research topics within three academic disciplines: biomedicine, physics, and engineering. The models were evaluated under three distinct conditions: zero-shot prompting, chain-of-thought prompting, and fine-tuning on existing ontologies. Additionally, we assessed the cross-discipline transferability of fine-tuned models by measuring their performance when trained in one discipline and subsequently applied to a different one. 
To support this analysis, we introduce \textit{PEM-Rel-8K}, a novel dataset consisting of over 8,000 relationships extracted from the most widely adopted taxonomies in the three disciplines considered in this study: MeSH, PhySH, and IEEE. Our experiments demonstrate that fine-tuning LLMs on \textit{PEM-Rel-8K} yields excellent performance across all disciplines.
\end{abstract}






\maketitle
\thispagestyle{empty}

\section{Introduction}

%Ontologies and taxonomies of research fields play an essential role in managing and organising scientific information, supporting effective  dissemination~\cite{Dunne2020,lipscomb2000medical,rous2012major}.

% \changed{test}

Ontologies and taxonomies of research fields are essential for structuring scientific knowledge, as they support the effective classification, dissemination, and retrieval of information~\cite{Dunne2020,lipscomb2000medical,rous2012major}. These knowledge bases are widely used to describe and categorise research outputs, including scientific articles, research projects, patents, datasets, and software. Moreover, they play a pivotal role in the evaluation and profiling of universities, research organisations, groups, and individual scholars~\cite{rahdari2021connecting,10.1162/qss_a_00021}. In addition, these ontologies are fundamental components of intelligent systems that operate on academic literature~\cite{osborne2013rexplore,beel2016paper,gusenbauer2020academic}, such as search engines~\cite{gusenbauer2020academic}, conversational agents~\cite{meloni2023integrating}, analytics dashboards~\cite{angioni2021aida}, and recommender systems~\cite{beel2016paper}.

%Finally, they are fundamental to intelligent systems which leverage academic literature~\cite{osborne2013rexplore,beel2016paper,gusenbauer2020academic}, such as search engines~\cite{gusenbauer2020academic}, conversational agents~\cite{meloni2023integrating}, analytics dashboards~\cite{angioni2021aida}, and recommender systems~\cite{beel2016paper}.

A recent survey identified 45 such knowledge organization systems (KOSs)—including ontologies, taxonomies, and thesauri—spanning most academic disciplines and highlighted several systemic issues within this landscape~\cite{salatino_survey_2024}. First, the coverage provided by these KOSs is highly uneven, with some disciplines either entirely absent or only partially represented. Second, these resources are weakly interconnected, which limits their capacity to effectively capture research activities that emerge from interdisciplinary work. Third, many of these KOSs are updated infrequently, failing to reflect recent developments, which often constitute the most dynamic and impactful aspects of scientific progress.
These shortcomings result in a limited representation of the scientific landscape, which in turn restricts the dissemination of research outputs and undermines the effectiveness of systems for literature analysis and exploration.

The root cause of many of these issues is that the creation and maintenance of such KOSs are expensive and time-consuming tasks, usually requiring the coordinated effort of multiple domain experts over a prolonged period~\cite{osborne2015klink}. In previous years, several approaches have been proposed to automate or semi-automate the ontology generation process~\cite{sanderson1999deriving,osborne2012mining,osborne2015klink,han2020wikicssh}. 
%Notable contributions in this direction include semi-automatic pipelines for generating OpenAlex Topics~\cite{openalex2024} and the fully automated development of the Computer Science Ontology~\cite{10.1162/dint_a_00055}, alongside other related initiatives. 
Despite these advancements, existing methods still encounter significant challenges in producing large-scale and fine-grained representations of research topics, due to the discipline’s inherent complexity and specialisation. Consequently, with a few notable exceptions, such as the Computer Science Ontology~\cite{10.1162/dint_a_00055}, most ontologies in this field are still developed predominantly through manual processes. 
%The automatic construction of comprehensive ontologies for research topics remains an open research problem.

Over the past three years, Large Language Models (LLMs) have significantly advanced the field of Natural Language Processing (NLP), enhancing the ability of machines to understand and generate human language. In this context, several studies have explored the use of LLMs for ontology generation, producing highly promising outcomes~\cite{babaei2023llms4ol,aggarwal2026large,fathallah2024neon,lippolis2025ontology}.

In this paper, we investigate the effectiveness of a broad range of LLMs and optimisation techniques in identifying semantic relations between pairs of research topics, a task that is central to the construction of academic ontologies. 
% In this paper, we investigate the effectiveness of a broad range of LLMs and optimisation techniques in identifying semantic relations between pairs of research topics, a task that has recently become central for developing new ontologies of academic fields.\color{black}
We focus on three strategies for adapting LLMs to this problem: zero-shot settings (ZSL), Chain-of-Thought (CoT) prompting, and fine-tuning on relationships drawn from existing ontologies. Our evaluation considers 12 open-weight LLMs, with sizes ranging from 3 billion to 27 billion parameters, and examines their ability to automatically detect three types of semantic relations: \textit{broader}, \textit{narrower}, and \textit{same-as}.

%In this paper, we evaluate the ability of a broad selection of LLMs and optimization techniques to identify semantic relations between pairs of research topics, a task that plays a key role in the construction of academic ontologies. Specifically, we assess the performance of 12 open-weight LLMs, ranging in size from 3 billion to 27 billion parameters, in automatically detecting three types of semantic relations: \textit{broader}, \textit{narrower}, and \textit{same-as}. 
%The models were evaluated under three conditions: zero-shot settings (ZSL), Chain-of-Thought (CoT) prompting, and fine-tuning on relationships from existing ontologies.

To support this analysis, we introduce \textit{PEM-Rel-8K}, a multi-disciplinary benchmark including over 8,000 relationships drawn from three widely used ontologies: \textit{IEEE}, which covers engineering and computer science; \textit{PhySH}, which focuses on physics; and \textit{MeSH}, the primary knowledge base used to categorise biomedical research. 
This novel resource also enabled us to explore the effects of fine-tuning a model on one scientific discipline and testing it on others, yielding valuable insights into the cross-discipline transferability of fine-tuning in this context.
To the best of our knowledge, this is the first large-scale study to investigate the use of LLMs for identifying semantic relationships between research topics across a wide range of scientific disciplines.

% We consider it a foundational step toward developing systems capable of automatically generating comprehensive and fine-grained representations of scientific knowledge across domains.

Our experiments demonstrate that fine-tuning LLMs on \textit{PEM-Rel-8K} yields excellent performance, significantly surpassing alternative approaches such as CoT prompting. Among the evaluated models, the fine-tuned \texttt{gemma-2-27b-it} achieved the highest F1-Score (93.5\%), followed by \texttt{gemma-2-9b-it} and \texttt{Mistral-Small-Instruct-2409}.
Notably, the best-performing LLM fine-tuned on \textit{PEM-Rel-8K} achieved an average F1-Score across the three disciplines that was only 1.2\% lower than that of models fine-tuned separately on discipline-specific training sets. Furthermore, our analysis of cross-discipline transferability shows that LLMs trained on one discipline can generalise effectively to others, maintaining competitive performance across different disciplines.

In summary, the contributions of this paper are as follows: \begin{itemize}
\item We present a comprehensive analysis of the capability of twelve LLMs to identify semantic relationships between research topics across three scientific disciplines:  engineering (IEEE), physics (PhySH), and biomedicine (MeSH).
\item We study different optimisation techniques, including ZSL, CoT reasoning, and fine-tuning.
\item We fine-tune the models on individual disciplines as well as on a hybrid training set that integrates all of them, enabling the analysis of cross-discipline adaptability.
\item We introduce and publicly release \textit{PEM-Rel-8K}, a novel modular benchmark designed for training and evaluating models on this task.
\item We provide the complete codebase for our analysis\footnote{The datasets and the code for our experiments are available at: \url{https://github.com/ImTanay/LLM-Multi-Domain-Ontology}}. \end{itemize}

The remainder of this paper is organised as follows. Section~\ref{sec:related-work} reviews the relevant literature. Section~\ref{sec:background} defines the task, introduces the novel \textit{PEM-Rel-8K} dataset, and describes the LLMs under analysis. Section~\ref{sec:methodology} outlines the experimental design and implementation setup. Section~\ref{sec:result} presents and discusses the results. Finally, Section~\ref{sec:conclusions} concludes the paper and suggests directions for future research.

\section{Related Work}\label{sec:related-work}

We begin by examining well-known ontologies of research areas (Section~\ref{sec:research-area-kos}), and then discuss various approaches for generating such ontologies (Section~\ref{sec:kos-generation}).

%This section reviews the relevant literature, starting with well-know ontologies of research areas (Section~\ref{sec:research-area-kos}) and approaches for their generation (Section~\ref{sec:kos-generation}).

\subsection{Knowledge Organization Systems of Research Areas}\label{sec:research-area-kos}

KOSs are formal frameworks designed to structure information and enable efficient knowledge management and retrieval~\cite{zeng2008knowledge, mazzocchi2018knowledge}. They are widely used to describe research areas and the relationships among them across a variety of repositories and digital libraries~\cite{salatino_survey_2024}. Depending on their structural complexity and available features, such as hierarchical depth or synonym control, these systems generally fall into four categories: term lists, taxonomies, thesauri, and ontologies.
A \textit{term list} is a flat, non-hierarchical collection of subject headings or descriptors used to index document sets, without explicit semantic relations among its entries~\cite{hedden2010taxonomies,Zaharee2013}.
A \textit{taxonomy} introduces structure by organising classes hierarchically through parent-child relationships~\cite{rasch1987nature}. It typically follows a tree structure, starting from a root node and branching into progressively more specific subclasses.
A \textit{thesaurus} extends the taxonomic model by incorporating additional descriptive properties, including definitions, associative links, and synonyms~\cite{niso2005}.
Finally, an \textit{ontology} provides a formal and explicit specification of a conceptual domain, classifying entities according to their defining attributes~\cite{gruber1993}. Ontologies represent the most functionally complete form of KOS, as they describe concepts, entities, and their relations~\cite{genesereth2012}. They support advanced semantic capabilities, including synonym resolution, property definition, and the representation of diverse relationship types~\cite{zeng2008knowledge}. Within the scholarly domain, these frameworks are essential for the classification and retrieval of research outputs, such as publications and datasets. 
The objective of this study is to assess the ability of LLMs to identify semantic relations, thereby supporting the construction and refinement of ontologies for the representation of research areas.

In this paper, we focus on three well-known academic KOSs: Medical Subject Headings, the IEEE Thesaurus, and PhySH~\cite{lipscomb2000medical,rous2012major,salatino2018computer,openalex2024}.
The Medical Subject Headings (MeSH) is an ontology that describes the field of medicine with over 31K headings~\cite{lipscomb2000medical}. It is maintained by the US National Library of Medicine, and it is primarily employed to organise medical publications in the Medline database.  
The IEEE Thesaurus includes approximately 12,000 curated terms relevant to electrical and electronics engineering. Developed and maintained by the IEEE, it plays a crucial role in classifying research outputs within IEEE’s digital repositories.
PhySH (Physics Subject Headings) offers about 3,700 concepts related to physics and is primarily used to index papers in Physical Review and on arXiv~\cite{smith2020physics}.

%The Computer Science Ontology (CSO) is a large-scale, automatically generated ontology in its field, encompassing 14K research concepts~\cite{salatino2018computer}, and it has been widely employed both in academia and industry. For instance, Springer Nature uses it to enhance publications metadata~\cite{salatino2019improving}, whereas Stanford University employed it to prepare its latest AI Index Report 2025~\cite{maslej2025artificialintelligenceindexreport}.

In addition to specialised ontologies, several multidisciplinary ones are available. These include the UNESCO Thesaurus and the ANZSRC Fields of Research (FoR), each comprising approximately 4,400 concepts; EuroVoc, which contains around 7,000; and OpenAlex Topics, which covers about 4,800 subjects. 

A recent survey~\cite{salatino_survey_2024} identified 45 ontologies of research topics, highlighting a highly fragmented landscape. % in which institutions often adopt incompatible representations. 
Notably, the survey found that no single ontology is openly available, fine-grained, and comprehensive across all research disciplines. This limitation is largely attributed to the reliance on manual curation, which requires significant time, expertise, and financial investment. 
Consequently, automated methods for generating research topic ontologies have gained increasing attention as a viable alternative. 
In this work, we contribute to this direction by proposing an approach for automatically identifying semantic relations between pairs of research topics.

\subsection{Automatic Generation of Ontologies}\label{sec:kos-generation}

Early semi-automatic approaches to ontology generation integrated expert input with linguistic and statistical tools~\cite{maedche2001learning}. With the advancement of NLP and machine learning techniques, these methods evolved into more sophisticated systems, such as Text2Onto~\cite{cimiano2005text2onto} and OntoLearn~\cite{navigli-etal-2004-quantitative}, which sought to minimise manual effort while maintaining a degree of human oversight. In recent years, the emergence of deep learning models such as BERT~\cite{devlin2018bert} has led to substantial improvements in concept extraction~\cite{grootendorst2022bertopic} and the modelling of hierarchical relationships~\cite{chen2020constructing,pisu2024leveraging}. %These advances include the use of hyperbolic embeddings to more effectively capture semantic hierarchies~\cite{le2019inferring}.
%Early semi-automatic approaches combined expert input with linguistic/statistical tools~\cite{maedche2001learning}, evolving with NLP and machine learning (e.g., Text2Onto~\cite{cimiano2005text2onto}, OntoLearn~\cite{navigli-etal-2004-quantitative}) to reduce manual effort, though validation remained key. Deep learning models, e.g., BERT~\cite{devlin2018bert}, later enhanced concept extraction (e.g., BERTopic~\cite{grootendorst2022bertopic}) and hierarchy modelling~\cite{chen2020constructing,pisu2024leveraging}, e.g., through hyperbolic embeddings~\cite{le2019inferring}.

LLMs have recently demonstrated significant potential in ontology generation. For instance, the LLMs4OL approach has achieved strong performance across a variety of datasets, including lexicosemantic knowledge, geographical information, and medical data~\cite{babaei2023llms4ol}. Nonetheless, several studies have revealed persistent challenges related to consistency and accuracy, underscoring the necessity of human oversight during the formalisation process~\cite{saeedizade2024navigating}.  
\cite{fathallah2024neon} introduced NeOn-GPT, a workflow that applies the NeOn methodology to ontology generation. %They findings suggest that LLMs are not yet fully capable of handling procedural tasks. However, recent advances in reasoning-oriented models may improve this limitation. 
Building on this work, LLMs4Life~\cite{fathallah2024llms4lifelargelanguagemodels} adapted NeOn-GPT for the development of life science ontologies. Their study emphasised the need for more context-rich prompts and demonstrated that the involvement of domain experts significantly enhances the quality of the results. In a related effort, \cite{lippolis2025ontology} assessed the use of LLMs in ontology design, aiming to reduce manual effort and support less experienced engineers. While their results confirmed the utility of LLMs as assistive tools, they also highlighted issues with consistency that require subsequent post-processing. In order to address these challenges, recent studies have proposed a range of human-in-the-loop strategies that investigate different modes of collaboration between LLMs and domain experts~\cite{tsaneva2025knowledge}.

Overall, while LLMs show promise for this task, they still face significant limitations in generating complete, high-quality ontologies that can match those created by domain experts, particularly in specialised domains such as the classification of scientific publications~\cite{sun2024large}. % This paper contributes to addressing these challenges by focusing on the identification of relationships between research topics.

The community focused on generating ontologies of research topics has developed several promising (semi-)automated methods. A notable example is Klink-2~\cite{osborne2015klink}, which enabled the construction of the Computer Science Ontology~\cite{salatino2018computer}. Klink-2 identifies both hierarchical and synonymous relationships by leveraging co-occurrence patterns, topic similarity, and subsumption logic. \cite{shen2018web} applied an extended subsumption technique to build Microsoft's Field of Science, an ontology containing over 200,000 concepts that was later integrated into the Microsoft Academic Graph. More recently, the OpenAlex team expanded the ASJC taxonomy by identifying more than 4,500 research topics using citation clustering, which they subsequently labelled using LLMs~\cite{openalex2024}. A similar methodology was independently adopted by~\cite{jenset2025large}, who developed a taxonomy of 29,000 research concepts and mapped it to the ANZSRC Fields of Research. 
Furthermore, ontology evaluation methods have been applied to enrich existing ontologies with additional research topics based on various requirements~\cite{kotis2020ontology,osborne2018pragmatic}. 

In conclusion, while LLMs have been employed to support a range of academic tasks, such as paper discovery~\cite{chow2024semantic}, citation prediction~\cite{buscaldi2024citation,hao2024hlm}, scientific question answering~\cite{auer2023sciqa,meloni2025exploring}, and literature review generation~\cite{bolanos2024artificial,scherbakov2024emergence}, their use for generating research topic ontologies has received limited attention. This study aims to fill this gap by systematically evaluating recent LLMs in terms of their ability to infer semantic relationships between research topics.

\section{Background}\label{sec:background}

This section formalises the task under investigation (Section~\ref{sec:task}), introduces the \textit{PEM-Rel-8K} dataset (Section~\ref{sec:meta-dataset}), and provides an overview of the LLMs used in our experiments (Section~\ref{sec:llms}).

%This section formalises the task under investigation (Section~\ref{sec:task}), describes the motivations and the use case (Section~\ref{sec:usecase}), introduces the PEM-Rel-8K dataset (Section~\ref{sec:meta-dataset}), and provides an overview of the LLMs used in our experiments (Section~\ref{sec:llms}).

% \newpage

\subsection{Task definition}\label{sec:task}

Identifying and formalising semantic relationships among research topics is essential for building academic KOSs, which play a key role in organising the scientific literature and improving information retrieval. These representations underpin many AI\footnote{AI - Artificial intelligence} systems for paper recommendation and research analysis, enable digital libraries and search engines to move toward robust semantic search, and support scientometric studies, for example, in the assessment of scientific impact and the forecasting of trends~\cite{salatino_survey_2024}.

In this paper, we address the task of identifying the semantic relationship between a given pair of research topics. More precisely, we formalise it as a single-label, multi-class classification problem, where each input pair of research topics, denoted as $t_A$ and $t_B$, is assigned to exactly one of the following mutually exclusive categories:
%
% The task addressed in this paper involves identifying the semantic relationship that exists between a given pair of research topics. We formalise this task as a single-label, multi-class classification problem, where each input pair of research topics, denoted as $t_A$ and $t_B$, is assigned to exactly one of the following mutually exclusive categories: 
\begin{itemize}
\setlength\itemsep{0cm}
\item \texttt{broader}: $t_A$ is a broader topic that subsumes the more specific topic $t_B$. For example, \textit{databases} subsumes \textit{distributed databases}. 
\item \texttt{narrower}: $t_A$ is a more specific topic subsumed by the broader topic $t_B$. For example, \textit{adaptive signal processing} is subsumed by \textit{signal processing}. This is the inverse relationship of \texttt{broader}.
\item \texttt{same-as}: $t_A$ and $t_B$ are semantically equivalent and can be used interchangeably across a broad range of information retrieval tasks. For example, \textit{ontology alignment} and \textit{ontology matching}.
\item \texttt{other}: in contrast with the previous three categories, this category does not define a semantic relation.  Its purpose is simply to provide the classifier with a mechanism to label negative examples. Without it, the classifier would be forced to assign one of the three predefined semantic relationships even when none actually applies to a given pair of topics.  
\end{itemize}

The first three relationships are widely employed in the construction of ontologies for research topics~\cite{smith2020physics}, as they are crucial for representing hierarchical structures as well as handling synonymous terms (otherwise known as synonym rings).
In practice, although we adopt simplified labels for these relations, they correspond directly to the standard Simple Knowledge Organization System (SKOS)\footnote{SKOS: Simple Knowledge Organization System --- \url{https://www.w3.org/TR/skos-reference}} properties \texttt{skos:narrower}, \texttt{skos:broader}, and \texttt{skos:exactMatch}. This design choice aligns with our primary objective of constructing research topic ontologies using SKOS as the underlying data model.
Conversely, as previously noted, the \texttt{other} category does not denote a formal semantic relationship and it simply provides a mechanism to avoid forced misclassification.

It is important to acknowledge that other semantic relationships, such as \textit{part-of}, \textit{instance-of}, \textit{has-attribute}, and \textit{has-process}, also play an important role in ontology development. However, this paper focuses on the core structural foundation of a research topics ontology, which is captured by the three relationships defined above.

\subsection{The PEM-Rel-8K dataset}\label{sec:meta-dataset}

% \color{blue}
The aim of this study is to evaluate model performance in a multidisciplinary context and to investigate the cross-discipline adaptability of models fine-tuned on data from a single discipline when applied to others. To this end, we sampled semantic relationships from three KOSs, each representing a distinct scientific discipline: the IEEE Thesaurus (Engineering), PhySH (Physics), and MeSH (Biomedicine).  \color{black}
We first constructed a dedicated dataset for each taxonomy, labelled \textit{IEEE-Rel-3K}, \textit{PhySH-Rel-875}, and \textit{MeSH-Rel-4K}, and then merged them into a unified benchmark named \textit{PEM-Rel-8K}.

The following sections detail the construction of each dataset and the subsequent integration into the final benchmark.

In the engineering discipline, we constructed \textit{IEEE-Rel-3K} by sampling 3,200 semantic relationships from the IEEE Thesaurus\footnote{IEEE Thesaurus -  \url{https://www.ieee.org/publications/services/thesaurus.html}}. Specifically, we used the latest available RDF version of the Thesaurus (v1.02)\footnote{IEEE Thesaurus (RDF) - \url{https://github.com/angelosalatino/ieee-taxonomy-thesaurus-rdf}}. We randomly selected 800 examples each for the \texttt{broader} and \texttt{narrower} relationships. 
Capturing instances of \texttt{same-as} was more challenging and required manual supervision, as the IEEE Thesaurus does not explicitly provide this relation. Instead, it relies on \texttt{skos:prefLabel} and \texttt{skos:altLabel} to indicate both related terms and synonyms. For example, ‘4G mobile communication’ is listed as a \texttt{skos:altLabel} of ‘5G mobile communication’. To address this, three experts manually analysed the set of topics connected through \texttt{skos:prefLabel} and \texttt{skos:altLabel}, and extracted 800 pairs of terms that were deemed to be true synonyms, lexical variants, or near-synonyms. These pairs were then annotated with the \texttt{same-as} relation.  
Finally, to produce the \texttt{other} pairs, we randomly generated 800 topic pairs that do not share any semantic link in the original ontology. %, and manually validated them to ensure their correctness.

In the physics discipline, we developed the \textit{PhySH-Rel-875} dataset by extracting 875 semantic relationships from the Physics Subject Headings\footnote{Physics Subject Headings (PhySH) - \url{https://github.com/physh-org/PhySH}}, which are provided in RDF format and adhere to the SKOS standard. From the April 2024 release, we randomly sampled 250 semantic relationships for each of the \texttt{broader} and \texttt{narrower} properties.
Similar to the IEEE taxonomy, PhySH does not explicitly include synonyms but uses the \texttt{skos:altLabel} property for alternative labels. As in the IEEE case, many topic pairs did not meet the criteria for the \texttt{same-as} relation. For instance, \textit{algebraic structure} is listed as an \texttt{skos:altLabel} of \textit{abstract algebra}. Three experts reviewed the 608 available \texttt{skos:altLabel} entries and were able to validate 125 instances as \texttt{same-as} relations. Finally, we generated 250 \texttt{other} pairs using the same procedure adopted for the IEEE dataset. This is the only imbalanced dataset in our collection. The imbalance is due to the difficulty of identifying a sufficient number of reliable \texttt{same-as} relations, combined with the intention to maintain a good sample size for the other relation types.

In the biomedical domain, we developed \textit{MeSH-Rel-4K}, a dataset comprising 4,000 semantic relationships extracted from Medical Subject Headings\footnote{Medical Subject Headings (MeSH) - \url{https://id.nlm.nih.gov/mesh/}}. The latest version of MeSH includes more than 31K subject headings described using a custom schema\footnote{MeSH Schema - \url{http://id.nlm.nih.gov/mesh/vocab}} (\texttt{mesh:}). 
In this schema, each topical subject heading is modelled as a \texttt{mesh:TopicalDescriptor}. 
These topics are connected by approximately 42K hierarchical relationships, expressed by \texttt{mesh:broaderDescriptor}. In addition, MeSH includes about 8K associative, non-hierarchical links between concept records, expressed by \texttt{mesh:relatedConcept}, which connect semantically related concepts.

We extracted the relationships for the \textit{MeSH-Rel-4K} dataset from the January 2025 release of MeSH. 
Specifically, we randomly sampled 1,000 \texttt{broader} and 1,000 \texttt{narrower} relationships, both derived from the \texttt{mesh:broaderDescriptor} property. We then sampled 1,000 \texttt{same-as} instances using the \texttt{mesh:relatedConcept} property. Finally, we manually curated 1,000 pairs of semantically unrelated topics to represent the \texttt{other} category.

Finally, we combined the three previously described datasets to construct \textit{PEM-Rel-8K}. This multi-disciplinary dataset comprises over 8,000 relationships spanning the three scientific disciplines. Each single-discipline dataset was divided into training, validation, and test sets following a 7:1:2 ratio. Table~\ref{tab:dataset-split} presents the sizes of both the individual and the combined datasets.

All datasets, along with the code used for their construction, are available at: \url{https://github.com/ImTanay/LLM-Multi-Domain-Ontology}.

\begin{table}[]
\centering
\caption{Sample distributions of \textit{PEM-Rel-8K}.\label{tab:dataset-split}}
\begin{tabular}{l|r|r|r|r|c}
\toprule
             & Train & Validation & Test  & Total & Percentage \\
\midrule
IEEE-Rel-3K  & 2,240 & 320        & 640   & 3,200 & 39.6 \%     \\
PhySH-Rel-87 & 613   & 87         & 175   & 875   & 10.8 \%     \\
MeSH-Rel-4K  & 2,800 & 400        & 800   & 4,000 & 49.6 \%      \\ \midrule
PEM-Rel-8K       & 5,653 & 807        & 1,615 & 8,075 &      \\
\bottomrule
\end{tabular}
\end{table}

\subsection{Large Language Models}\label{sec:llms}

In this study, we evaluated twelve decoder-only LLMs. 
Table~\ref{table:llms} presents an overview of the LLMs, listing the model names, the shorter aliases used throughout this paper, the number of parameters, the context window sizes, and the Low-Rank Adaptation (LoRA)~\cite{hu2021lora} parameters employed for their fine-tuning: \begin{enumerate*}[label=\roman*)] \item \textbf{r}, the rank of the update matrices (lower values produce smaller update matrices with fewer trainable parameters), and \item \textbf{alpha}, the LoRA scaling factor. \end{enumerate*}
The selected models vary in size and represent several prominent families: Mistral~\cite{jiang2023mistral} (three models), Llama~\cite{llama3modelcard} (three models), Gemma~\cite{team2024gemma} (three models), Phi~\cite{abdin2024phi3technicalreporthighly,abdin2024phi4technicalreport} (two models), and Zephyr~\cite{huggingfaceUnslothzephyrsftHugging} (one model). The number of parameters ranges from 2.51 billion (\texttt{gemma-2b}, the smallest) to 27.2 billion (\texttt{gemma-27b}, the largest). All models were quantised to 4-bit precision and are publicly accessible via HuggingFace.

%Table~\ref{table:llms} provides an overview of the LLMs assessed, reporting the model name, context window size, and Low-Rank Adaptation (LoRA)~\cite{hu2021lora} parameters: 
%\begin{enumerate*}[label=\roman*)] 
 %   \item \textbf{r}: the rank of the update matrices, expressed as an integer; a lower rank results in smaller update matrices with fewer trainable parameters, 
 %   \item \textbf{alpha}: the LoRA scaling factor. 
%\end{enumerate*}

%, specified as an integer
\input{tables/llms}

\section{Experimental Methodology}\label{sec:methodology}

We conducted an extensive series of experiments to evaluate the performance of LLMs on the \textit{PEM-Rel-8K} benchmark.  
A distinguishing characteristic of this benchmark is its modular structure, which enables training on each of the three discipline-specific datasets individually (\textit{IEEE-Rel-3K}, \textit{MeSH-Rel-4K}, and \textit{PhySH-Rel-875}) as well as on their combined form. Testing can likewise be performed on each of the individual datasets or on the aggregate. %This design supports a detailed investigation into which training configurations yield the best results for specific testing scenarios.

We explored two main evaluation settings: zero-shot learning and fine-tuning.

In the zero-shot learning setting, we examined model performance using two different prompting strategies. Evaluations were carried out on the multidisciplinary test set as well as on each of the three discipline-specific test sets.
% Altogether, the combination of two strategies, 12 LLMs (Table~\ref{table:llms}), and four test set datasets (Table~\ref{tab:dataset-split}) produced 96 experiments.

In the fine-tuning setting, we systematically assessed three conceptually distinct scenarios:

\begin{enumerate}
\item \textbf{Discipline-specific evaluation:} The LLM is fine-tuned and tested within the same discipline (e.g., fine-tuned on the training set of \textit{MeSH-Rel-4K} and evaluated on the test set of \textit{MeSH-Rel-4K}), a setting expected to yield the highest performance due to discipline alignment.
\item \textbf{Cross-discipline evaluation:} The LLM is fine-tuned on one discipline and tested on another (e.g., fine-tuned on the training set of \textit{MeSH-Rel-4K} and evaluated on the test set of \textit{PhySH-Rel-875}) as well as on the entire \textit{PEM-Rel-8K} to assess cross-discipline transferability.
\item \textbf{Multidisciplinary evaluation:} The LLM is fine-tuned using the training set of \textit{PEM-Rel-8K} and evaluated on its corresponding test set as well as on the test sets of three discipline-specific datasets to assess its robustness across different disciplines.
\end{enumerate}

The resulting experimental framework includes 24 distinct combinations of train and test sets. Each was applied across 12 different LLMs, resulting in a total of 288 experimental runs. 
%
% \changed{Table~\ref{tab:dataset-split} reports the sizes and splitting configuration of the datasets used for these experiments.}
%
Table~\ref{tab:dataset-split} summarises the sizes of the datasets used in these experiments and their partitioning.

We evaluated the performance of all models using macro-averaged precision, recall, and F1-score, as these are standard metrics for classification tasks.

In the next two subsections, we provide a detailed description of the procedures used in both the zero-shot and fine-tuning experiments. 
Additional technical specifications, including the libraries and hardware used, are provided in Section~\ref{sec:experimental-setup}.

%\color{red}
%To evaluate the twelve LLMs on our datasets, we conducted three series of experiments:  
%\begin{enumerate*}[label=\roman*)]  
%    \item zero-shot learning using various prompting strategies,  
%    \item fine-tuning on each dataset, and  
%    \item transfer learning, in which models fine-tuned on one dataset were evaluated on a different domain-specific test set.  
%\end{enumerate*}  
%These are detailed in the following subsections, with the experimental setup presented in Section~\ref{sec:experimental-setup}.
%\color{black}

\subsection{Zero-Shot Prompting Strategies}\label{sec:prompting}

We implemented two prompting strategies: standard prompting and bidirectional CoT prompting~\cite{aggarwal2026large}.

The standard prompting, employed as baseline, generates prompts for each pair of research topics via a predefined template (available in the GitHub\footnote{Our prompts - \url{https://github.com/ImTanay/LLM-Multi-Domain-Ontology}} repository). This template outlines the task, defines the four relationships, and mandates a specific format for the output to facilitate parsing. To ensure a fair comparison, we employed the same prompt for all models.

Bidirectional CoT prompting, introduced by \cite{aggarwal2026large}, was included in this study due to its strong performance in classifying semantic relationships between research topics. This technique builds upon CoT prompting, which has shown effectiveness across a broad range of complex tasks~\cite{wei2022chain,kojima2023largelanguagemodelszeroshot}. 
The approach involves two sequential prompts. The first prompt asks the model to provide definitions for both topics, construct a sentence incorporating them, and reflect on their potential semantic relationship. The second prompt uses the response generated by the first, adds instructions for the classification task, and presents it back to the model.
This process is repeated with the positions of the two topics swapped to achieve bidirectionality. Finally, a rule-based referee is applied to resolve any discrepancies between the outcomes of the two runs. All prompt templates and referee rules follow those described in the original paper~\cite{aggarwal2026large}.

In total, the combination of two strategies, 12 LLMs (Table~\ref{table:llms}), and four datasets (Table~\ref{tab:dataset-split}) resulted in 96 experiments.

\subsection{Fine-Tuning}\label{sec:fine-tuning}

The fine-tuned models were obtained by training the 12 LLMs on the four datasets: \textit{PEM-Rel-8K}, \textit{IEEE-Rel-3K}, \textit{MeSH-Rel-4K}, and \textit{PhySH-Rel-875}. This process yielded a total of 48 fine-tuned models. 
To ensure compatibility and comparability of results, we consistently used the same training and validation sets (split 7:1:2, as detailed in Section~\ref{sec:meta-dataset}) and employed a uniform prompt-based fine-tuning approach. This involved formatting each pair of research topics into a conversational prompt using the following predefined template: %, kept identical across all experiments:
\begin{Verbatim}[fontsize=\small,frame=single]
Classify the relationship between `[TOPIC-A]' and `[TOPIC-B]'
\end{Verbatim}

In this format, the placeholders \texttt{`[TOPIC-A]'} and \texttt{`[TOPIC-B]'} were dynamically replaced with the corresponding surface forms of topics $t_A$ and $t_B$, as follows:
% \color{red}
% \begin{Verbatim}[fontsize=\small,frame=single]
% user: Classify the relationship between `Coal industry' and `Coal sector'
% model: relationship: exactMatch
% \end{Verbatim}
% \color{black}
\begin{Verbatim}[fontsize=\small,frame=single]
user: Classify the relationship between `Biology' and `Genetics'
model: relationship: broader
\end{Verbatim}
% In the training and validation sets, we explicitly included the label in the format: \texttt{relationship: [`relationship-type']} to encourage consistent model outputs, facilitating easier parsing of the predicted relationships. Our parser maps these simplified labels to corresponding \texttt{skos} predicates: \texttt{broader} to \texttt{skos:broader}, \texttt{narrower} to \texttt{skos:narrower}, \texttt{same-as} to \texttt{skos:exactMatch}, and \texttt{other} to \texttt{skos\_ext:other}. 

During training and validation, we explicitly included the expected output in the format (\texttt{relationship: [RELATIONSHIP-TYPE]}) to guide the model towards producing structured and easily parsable responses. This design ensured consistency and simplified the extraction of predicted labels. 
The resulting 48 models were then tested on the four datasets to examine all the evaluation scenarios discussed previously, constituting 192 distinct experiments.

\subsection{Experimental Setup}\label{sec:experimental-setup}

To facilitate the fine-tuning and interaction with the set of LLMs reported in Table~\ref{table:llms}, we employed two open-source libraries: KoboldAI\footnote{KoboldAI - \label{kobold}\url{https://github.com/KoboldAI/KoboldAI-Client}} and Unsloth~\cite{unsloth}.

\textbf{KoboldAI}, is an open-source platform built on top of llama.cpp\footnote{llama.cpp - \url{https://github.com/ggml-org/llama.cpp}} that provides API-based access to LLMs hosted locally. We used KoboldAI to interact with the 12 models in a zero-shot setting.%, employing various prompting strategies as detailed in Section~\ref{sec:prompting}. This setup enabled controlled evaluation of model behaviour without fine-tuning.

\textbf{Unsloth} was employed for fine-tuning and transfer learning experiments. It is an open-source Python library built on top of the Hugging Face Transformers library~\cite{wolf-etal-2020-transformers} and PEFT (Parameter-Efficient Fine-Tuning) method~\cite{peft}. Unsloth supports 4-bit quantised training via BitsAndBytes\footnote{BitsAndBytes - \url{https://github.com/bitsandbytes-foundation/bitsandbytes}}, which significantly reduces memory usage and enables training on consumer-grade GPUs. It also integrates LoRA~\cite{hu2021lora}, allowing parameter-efficient fine-tuning by injecting lightweight trainable adapter layers into frozen pre-trained models. %Additionally, Unsloth offers a simplified interface for converting plain-text or prompt-based data into model-ready conversational formats with minimal code overhead, making it well-suited for fine-tuning on prompt-driven datasets.

All experiments were conducted on Google Colaboratory instances equipped with NVIDIA A100 and L4 GPUs. 
To ensure transparency and reproducibility, the complete codebase is available in the GitHub repository\footnote{Our code – \url{https://github.com/ImTanay/LLM-Multi-Domain-Ontology}}. It includes all scripts for prompting and fine-tuning, along with detailed configuration parameters for both Unsloth and KoboldAI.

\input{tables/transfer-table}

\section{Results}\label{sec:result} 
In this section, we present and discuss the experimental results. % following the same order as in Section 4. 
We begin with zero-shot learning, followed by the three evaluation scenarios involving fine-tuning. Finally, we compare different models by considering the most comprehensive case, in which they were both fine-tuned and evaluated on the full \textit{PEM-Rel-8K} dataset.

%We conducted a stunning total of 288 experiments by training 12 LLMs across 24 distinct combinations of experimental conditions, encompassing three major approaches: 
%\begin{enumerate*}[label=\roman*)]
%    \item evaluation of LLMs using two prompting strategies (8 experimental combinations and 96 experiments),
%    \item fine-tuning each model on its respective domain-specific dataset and evaluating on the same dataset (7 experimental combinations and 84 experiments), and
%    \item transfer learning experiments, where models fine-tuned on one dataset were evaluated on alternative test sets (9 experimental combinations and 108 experiments).
%\end{enumerate*}

Table~\ref{table:transfer-table} reports the best-performing models and their results across the 24 experimental combinations. 
Among the 12 models evaluated, only five achieved top performance: \texttt{gemma-27b}, which ranked first in eight cases; \texttt{mistral-22b} and \texttt{phi-4}, each leading in five cases; and \texttt{mistral-7b} and \texttt{gemma-9b}, each with three top results.
%Among the 12 models evaluated, only five appear here: \texttt{gemma-27b} (the best in eight cases), \texttt{mistral-small} and \texttt{phi-4} (five each), and \texttt{mistral} and \texttt{gemma-9b} (three each).

\subsection{Zero-shot Strategies}\label{results:prompting}

Bidirectional CoT consistently outperformed standard prompting across all evaluated models and datasets, in agreement with previous findings~\cite{wei2022chain,aggarwal2026large}. Specifically, the CoT approach resulted in an average F1-score improvement of 5.1\% (standard deviation $\pm 0.3$\%) compared to simple prompting across the four datasets. When applied to the full datasets, CoT achieved a robust F1 score of 73.0\%, exceeding the performance of simple prompting by 5.3 percentage points. 
%These results indicate that, while LLMs can achieve reasonable performance in a zero-shot setting, employing advanced prompting strategies remains essential for enhancing their effectiveness.

In the experiment using simple prompting, \texttt{mistral-22b} and \texttt{gemma-9b} achieved the highest performance. However, when applying bidirectional CoT, the smaller \texttt{mistral-7b} outperformed the other models on three out of four datasets (\textit{IEEE-Rel-3K}, \textit{PhySH-Rel-875}, and \textit{PEM-Rel-8K}). These results underscore the importance of structured prompting techniques, particularly when deploying smaller and more cost-efficient models.

\subsection{Fine-Tuning}\label{results:fine-tuning}
In the following, we discuss the three evaluation scenarios presented in Section~4: discipline-specific evaluation, cross-discipline evaluation, and multidisciplinary evaluation.

\subsubsection{Discipline-specific evaluation.}

The highest performance on each of the three discipline-specific test sets was achieved by models fine-tuned on the corresponding training data. However, performance varied substantially across the test sets, and the top-performing model was different for each one. Specifically, \texttt{gemma-27b} achieved the highest overall F1-score on \textit{IEEE-Rel-3K} (98.9\% F1), while \texttt{gemma-9b} delivered the best performance on \textit{PhySH-Rel-875} (93.6\%), and \texttt{phi-4} performed best on \textit{MeSH-Rel-4K} (91.7\%).

The excellent results on \textit{IEEE-Rel-3K} may be attributed to the nature of the relevant topics in electrical engineering and computer science, which are likely to be well represented in the LLM's training data, such as code repositories, technical documents, manuals, and patents. Notably, fine-tuning on the much smaller \textit{PhySH-Rel-875} still produced a high F1-score, demonstrating the effectiveness of fine-tuning even when using a relatively small training set.

%The near-perfect results achieved when fine-tuning on \textit{IEEE-Rel-3K} are likely attributable to the nature of its training data, which probably includes extensive code repositories, technical documents, manuals, and patents. Crucially, fine-tuning on a significantly smaller dataset like \textit{PhySH-Rel-875} also yielded high performance (an F1-score of 0.936 with gemma-9b), suggesting that this approach remains highly beneficial even in low-resource scenarios. Instead, although outstanding, the lowest results in fine-tuning \textit{MeSH-Rel-4K}, are attributed to the complexity of correctly classifying semantic relationships like \texttt{skos:broader} and \texttt{skos:exactMatch}.

\subsubsection{Cross-discipline evaluation.}

The cross-discipline experiments demonstrate the generalisation capabilities of LLMs when fine-tuned on one discipline and evaluated on another. This behaviour is analogous to transfer learning, where knowledge acquired from solving one task or operating in one domain is applied to enhance performance on different but related tasks~\cite{zhuang2020comprehensivesurveytransferlearning}. 

Fine-tuning models on any single discipline resulted in highly competitive performance. In particular, for each discipline, the best cross-discipline model achieved F1 scores that were, on average, only 5.1\% (standard deviation $\pm 1.7$\%) F1 points lower than those of the model trained directly on the same discipline. Moreover, these models provided a substantial average improvement of 16.1 percentage points (standard deviation $\pm 1.6$\%) over the best zero-shot strategy.

The optimal cross-discipline training sets and models varied depending on the test set. When evaluating on \textit{IEEE-Rel-3K}, the best-performing solution was \texttt{phi-4} fine-tuned on \textit{PhySH-Rel-875} (94.7\% F1). For \textit{MeSH-Rel-4K}, the top model was \texttt{gemma-27b} fine-tuned on \textit{PhySH-Rel-875} (87.7\% F1). Finally, for the \textit{PhySH-Rel-875} test set, the best-performing model was \texttt{phi-4} fine-tuned on \textit{MeSH-Rel-4K} (86.4\% F1).

The results confirm that models fine-tuned on certain STEM disciplines can be successfully applied to others, suggesting that it is not necessary to use discipline-specific datasets to achieve solid performance in this task. 
However, these findings should be further validated across disciplines that are typically classified into distinct domains, such as physics and the social sciences.

%Moreover, smaller models, such as \texttt{phi-4}, demonstrate remarkable adaptability in this context.

\subsubsection{Multidisciplinary evaluation. }
The purpose of this evaluation is to determine whether models trained on the multidisciplinary \textit{PEM-Rel-8K} dataset yield consistently strong results across different disciplines. The results support this hypothesis, as models trained on the full dataset performed excellently on the three discipline-specific test sets. Their F1-scores were only 1.2\% lower than those of models trained exclusively on discipline-specific datasets. In all cases, the best-performing model was \texttt{gemma-27b}, which achieved F1-scores of 97.3\% on \textit{IEEE-Rel-3K}, 90.8\% on \textit{MeSH-Rel-4K}, and 92.5\% on \textit{PhySH-Rel-875}.

These findings confirm that \textit{PEM-Rel-8K} can be used to develop robust models that generalise well across multiple disciplines and are suitable for producing multidisciplinary ontologies.

When considering the full \textit{PEM-Rel-8K} test set, the best solution was again \texttt{gemma-27b} fine-tuned on the same dataset, achieving an F1-score of 93.5\%. Notably, \texttt{phi-4}, trained on \textit{MeSH-Rel-4K}, also produced competitive results with an F1-score of 91.9\%. This further reinforces the conclusions of the cross-discipline evaluation.

%On \textit{PEM-Rel-8K}, gemma-27b performed best with 0.935 F1-Score.

%On the other hand, models trained on the multi-domain \textit{PEM-Rel-8K} consistently ranked as the second-best performers on all test sets, only outdone by models trained directly on the target domain, showcasing \textit{PEM-Rel-8K}'s value as a transfer learning source.
%Notably, gemma-27b, when fine-tuned on \textit{PEM-Rel-8K}, demonstrated excellent generalisation capabilities, achieving F1-scores of 0.973 on \textit{IEEE-Rel-3K}, 0.908 on \textit{MeSH-Rel-4K}, and 0.925 on \textit{PhySH-Rel-875}. This suggests a strong ability to internalise semantic features applicable across disciplinary boundaries.

\subsection{Comparing LLMs on PEM-Rel-8K}\label{sec:comparative-analysis}
 
As observed in previous analyses, fine-tuning on \textit{PEM-Rel-8K} yielded strong results across all disciplines. This section presents a more detailed investigation of the LLMs that were fine-tuned and evaluated using the full \textit{PEM-Rel-8K} dataset.

Table~\ref{table:pem-vs-pem} reports the precision, recall, and F1-score for all evaluated models. %It also provides a breakdown of these metrics for the four categories. 
It also provides a breakdown of these metrics across the four categories: the three relations and the \texttt{other} category that captures cases in which the two topics are not connected by any of the defined relations.

In line with the broader findings presented in Table~\ref{table:transfer-table}, \texttt{gemma-27b} emerges as the best-performing model overall, achieving a F1-score of 93.5\%. Several other models also demonstrate strong performance on this benchmark, notably \texttt{gemma-9b} (92.6\% F1), \texttt{mistral-22b} (92.2\% F1), and \texttt{phi-4} (91.8\% F1).

\input{tables/pem-vs-pem}

Overall, \texttt{gemma-27b} achieves the highest F1-score across all relation types and also obtains the highest recall for \texttt{narrower} (92.4\%). However, some models demonstrate notable strengths in specific relations. For example, \texttt{mistral-22b} achieves the highest precision for \texttt{broader} (96.3\%) and the best recall for \texttt{same-as} (95.3\%). \texttt{gemma-2b} attains the highest precision for \texttt{narrower} (94.3\%), while \texttt{gemma-9b} excels in precision for \texttt{same-as} (90.9\%) and recall for \texttt{other} (97.3\%). Furthermore, \texttt{phi-4} obtains the best precision for \texttt{other} (97.0\%) and achieves the highest recall for \texttt{broader} (92.7\%).

The four categories achieve comparable average F1-scores across the models, with values ranging from 88\% to 94\%. However, \texttt{other} consistently achieves the highest average scores across all metrics (F1-score: 94.3\%, precision: 94.6\%, and recall: 94.2\%). Although \texttt{other} is not a formal semantic relation, as discussed in Section~\ref{sec:task}, these strong results indicate that LLM-based systems are very effective at identifying cases in which pairs of topics are not connected by any of the predefined relations. This capability is particularly valuable, as incorrectly identified relations, especially hierarchical ones, can compromise the quality of the resulting ontology by introducing cycles or enabling incorrect inferences~\cite{osborne2015klink}.

%In contrast, \texttt{narrower} shows difficulties in recall, with the lowest average value of 87.7\%. 

Notably, \texttt{same-as} proves the most challenging relation, recording the lowest F1-score (average 88.5\%) and precision (average 87.2\%). This relatively lower performance can be attributed to the inherent difficulty of defining \texttt{same-as} with precision, as even different ontologies and thesauri across various disciplines often interpret and apply this relation inconsistently.

% \changed{\subsection{Analysis of Misclassification}}

\subsection{Analysis of Best-Performing LLMs on PEM-Rel-8K}

\begin{figure*}[t!]
    \centering
    \begin{subfigure}[t]{0.50\textwidth}
        \centering
        \includegraphics[width=\linewidth]{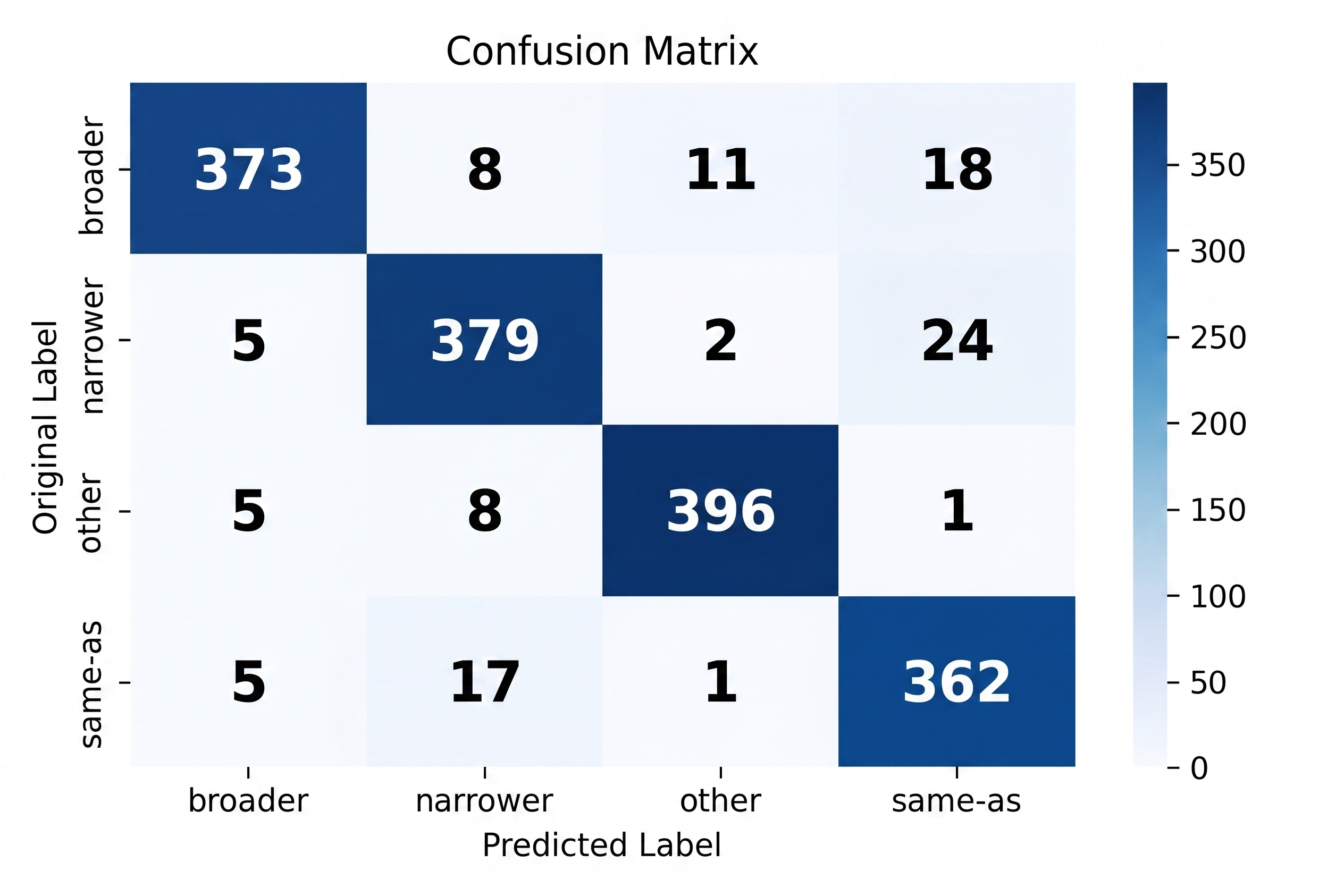}
        \caption{\texttt{gemma-27b} (F1-Score: 93.50\%)}\label{gemma-27b}
    \end{subfigure}%
    \begin{subfigure}[t]{0.50\textwidth}
        \centering
        \includegraphics[width=\linewidth]{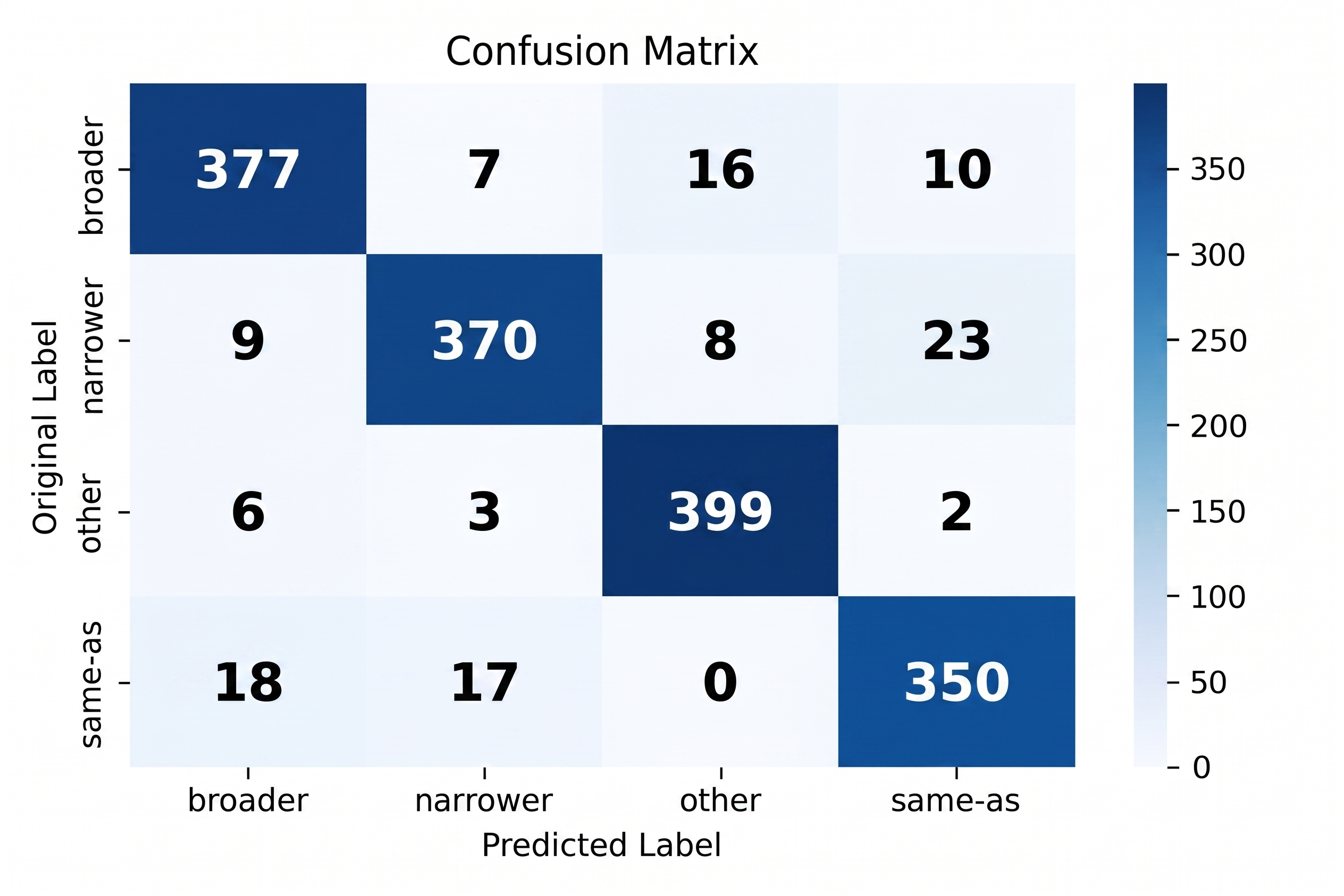}
        \caption{\texttt{gemma-9b} (F1-Score: 92.60\%)}\label{gemma-9b}
    \end{subfigure}
    \begin{subfigure}[t]{0.50\textwidth}
        \centering
        \includegraphics[width=\linewidth]{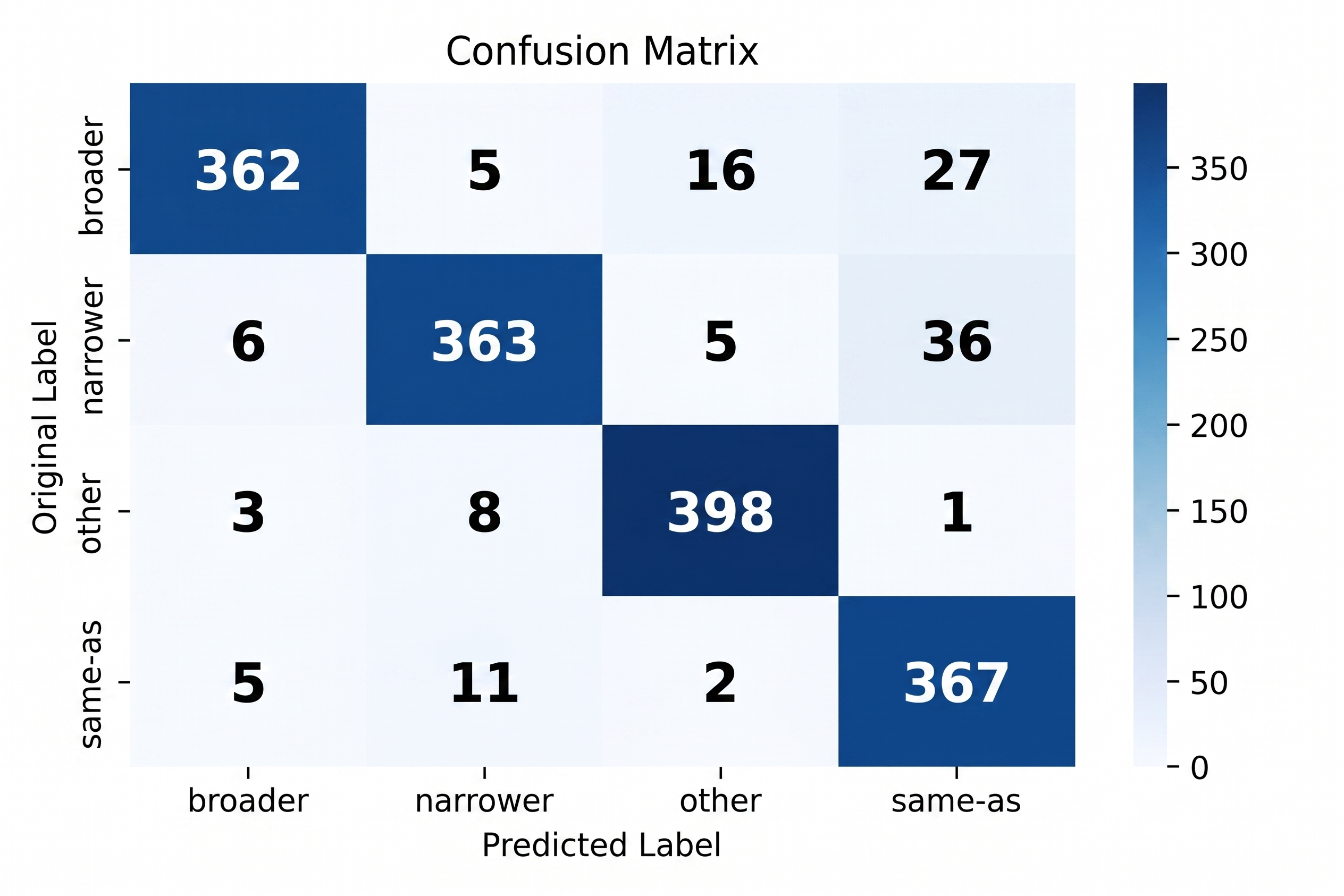}
        \caption{\texttt{mistral-22b} (F1-Score: 92.20\%) }\label{mistral-22b}
    \end{subfigure}%
    \begin{subfigure}[t]{0.50\textwidth}
        \centering
        \includegraphics[width=\linewidth]{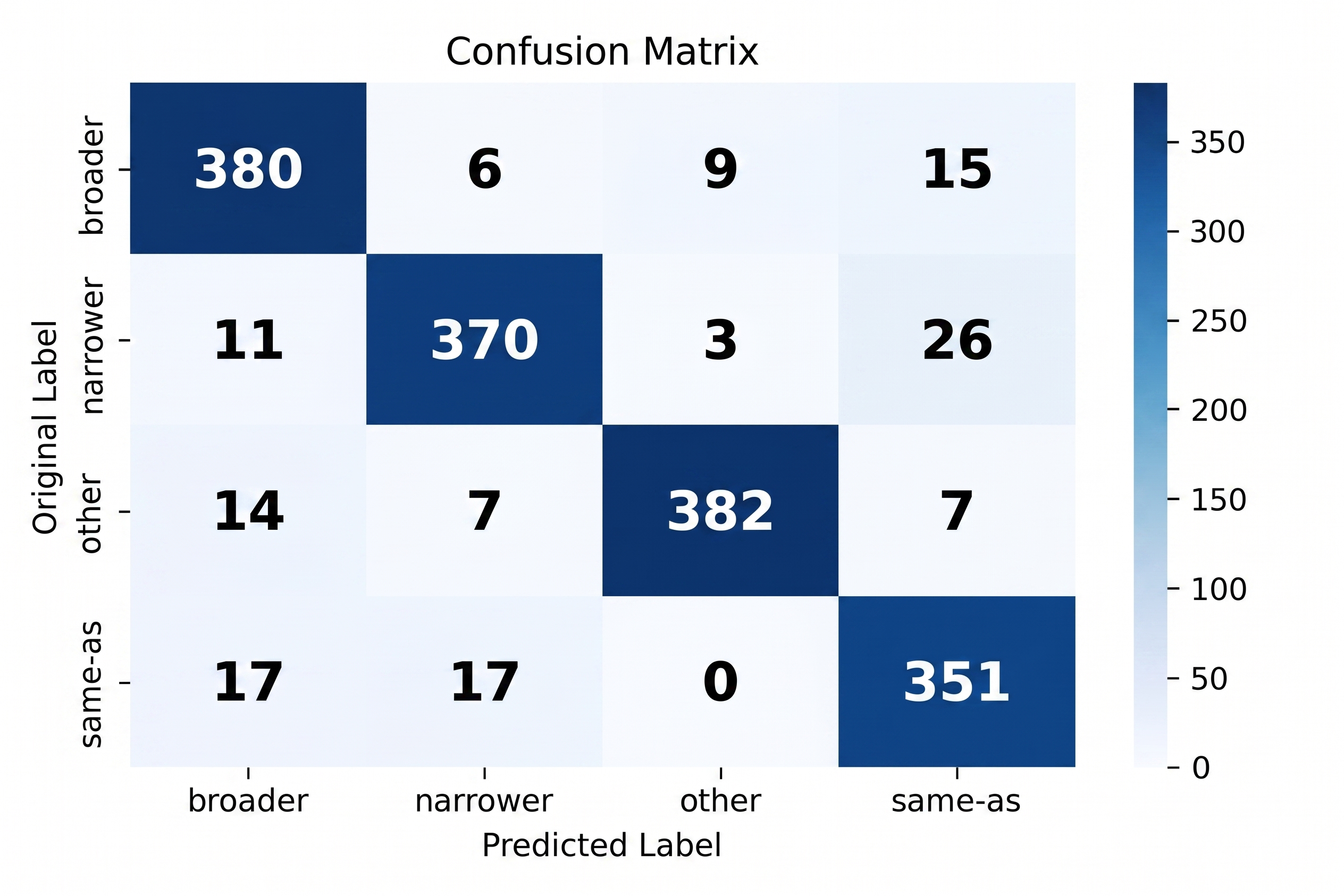}
        \caption{\texttt{phi-4} (F1-Score: 91.80\%)}\label{phi-4}
    \end{subfigure}
    \begin{subfigure}[t]{0.50\textwidth}
        \centering
        \includegraphics[width=\linewidth]{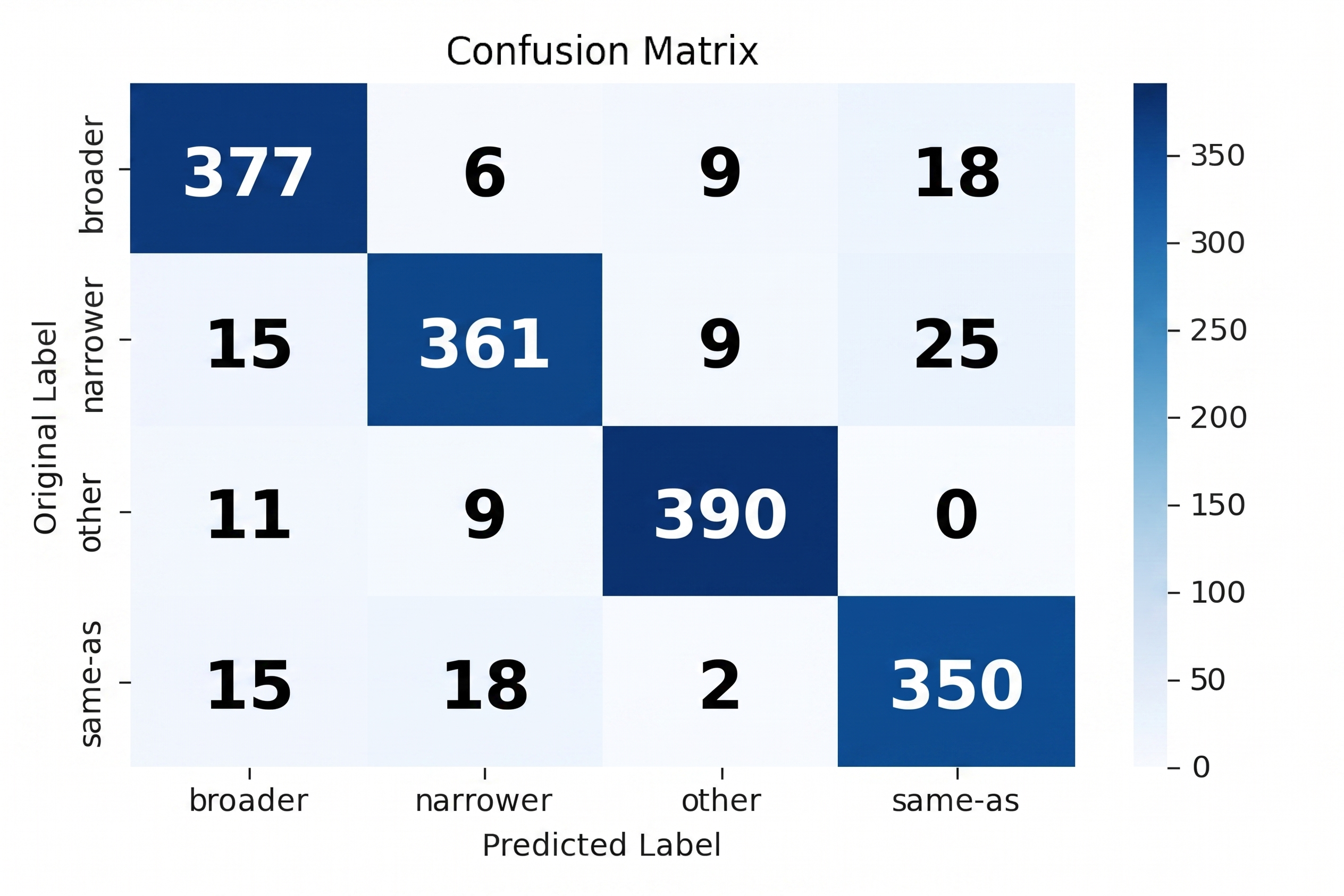}
        \caption{\texttt{zephyr-7b} (F1-Score: 91.50\%)}\label{zephyr-7b}
    \end{subfigure}
    \caption{Confusion matrices for best-performing LLMs on \textit{PEM-Rel-8K}}
    \label{fig:confusionmatrices}
\end{figure*}

The confusion matrices in Figure~\ref{fig:confusionmatrices} provide a comprehensive evaluation of the performance of the top 5 LLMs on the \textit{PEM-Rel-8K}. These models (\texttt{gemma-27b}, \texttt{gemma-9b}, \texttt{mistral-22b}, \texttt{phi-4}, and \texttt{zephyr-7b}) achieved high overall F1-scores, ranging from 93.5\% to 91.5\%, with \texttt{gemma-27b} performing the best.

All five models exhibit strong diagonal dominance in their respective confusion matrices, reflecting high classification accuracy across all four relation types: \texttt{broader}, \texttt{narrower}, \texttt{other}, and \texttt{same-as}. Among these, the \texttt{other} category consistently demonstrates the highest classification accuracy, with correct predictions exceeding 380 (out of 410) instances across all models. This suggests that distinguishing non-hierarchical, non-equivalent relations is comparatively easier. 
%due to their more distinct lexical patterns.

Across all models, the most common error involves misclassifying hierarchical relations (\texttt{broader}/\texttt{narrower}) as equivalence (\texttt{same-as}). The opposite error, namely misclassifying \texttt{same-as} relations as hierarchical, occurs less frequently but remains noticeable.
In details, \texttt{gemma-27b} (F1: 93.5\%, see Fig.~\ref{gemma-27b}) exhibits best performance, particularly for the \texttt{narrower} and \texttt{other}, correctly classifying 379 and 396 instances, respectively. Misclassifications are minimal, although some confusion persists between \texttt{same-as} and the hierarchical categories. Specifically, 18 instances of \texttt{broader} were misclassified as \texttt{same-as}, while 5 instances of \texttt{same-as} were incorrectly labelled as \texttt{broader}. Similarly, 24 instances of \texttt{narrower} were predicted as \texttt{same-as}, and 17 \texttt{same-as} instances were misclassified as \texttt{narrower}. \texttt{gemma-9b} (F1: 92.6\%, see Fig.~\ref{gemma-9b}), the smaller variant of \texttt{gemma-27b}, demonstrates slightly reduced performance, particularly in distinguishing \texttt{same-as} from hierarchical relations. Notably, 18 \texttt{same-as} instances were misclassified as \texttt{broader} and 17 as \texttt{narrower}, while 10 \texttt{broader} and 23 \texttt{narrower} instances were predicted as \texttt{same-as}. %This pattern indicates a diminished capacity to accurately identify equivalence relations. 
\texttt{mistral-22b} (F1: 92.2\%, see Fig.~\ref{mistral-22b}) performs comparably to the two \texttt{gemma} models but exhibits increased confusion between \texttt{same-as} and other relation types. In particular, 36 instances of \texttt{narrower} were misclassified as \texttt{same-as}, alongside 27 \texttt{broader} instances misclassified as \texttt{same-as}, and 16 as \texttt{other}. \texttt{phi-4} (F1: 91.8\%, see Fig.~\ref{phi-4}) also displays considerable confusion between \texttt{same-as} and hierarchical relations. Specifically, 26 \texttt{narrower} and 15 \texttt{broader} instances were misclassified as \texttt{same-as}, while 34 \texttt{same-as} instances were incorrectly labelled as \texttt{broader} (17) and \texttt{narrower} (17), reflecting a bidirectional confusion pattern. \texttt{zephyr-7b} (F1: 91.5\%, see Fig.~\ref{zephyr-7b}), the smallest among the top five models, exhibits the highest rate of misclassification. It frequently confuses \texttt{narrower} and \texttt{same-as} (25 and 18 misclassifications, respectively), as well as \texttt{broader} and \texttt{same-as} (18 and 15 misclassifications, respectively), indicating a greater difficulty in disentangling equivalence from hierarchical semantics.

% An analysis of the data revealed that the most frequent error pattern, namely the misclassification of the hierarchical relation as \texttt{same-as}, is largely due to the presence of overlapping lexical cues and the contextual ambiguity that often characterises the natural language expressions of these relations across different ontologies. 
An analysis of the data revealed that the most frequent error pattern, namely the misclassification of the hierarchical relation as \texttt{same-as}, is largely due to lexical overlap between terms and inconsistent definitions of semantic relationships across different ontologies. 
% For example, in MeSH, the term ``microsporidians'' is defined as a broader concept of ``microsporea''; however, all of the best-performing models (\texttt{gemma-27b}, \texttt{gemma-9b}, \texttt{mistral-22b}, \texttt{phi-4}, and \texttt{zephyr-7b}) predicts this pair as \texttt{same-as}.
Consider, for instance, the MeSH classification: ``microsporea'', a class of fungi, is a more specific term than ``microsporidians'', a group of spore-forming unicellular parasites. Despite this, every top-performing model (\texttt{gemma-27b}, \texttt{gemma-9b}, \texttt{mistral-22b}, \texttt{phi-4}, and \texttt{zephyr-7b}) misidentified them as being \texttt{same-as}.
Another similar example comes from the IEEE Thesaurus, where ``Nanoscale technology'' is a more specific concept than ``Nanotechnology''. Yet, the five models all considered these two topics to be the same.

\section{Conclusions}\label{sec:conclusions} 

In this paper, we have presented a comprehensive analysis of the performance of a diverse set of LLMs in identifying semantic relations between pairs of research topics. This task is essential for the construction of ontologies structuring research fields, which are critical for managing and organising scientific knowledge~\cite{salatino_survey_2024}. 

To support this analysis, we have introduced \textit{PEM-Rel-8K}, a multidisciplinary and modular benchmark created by aggregating three datasets extracted from \textit{IEEE}, \textit{PhySH}, and \textit{MeSH}. 
Our experiments show that fine-tuning LLMs on \textit{PEM-Rel-8K} yields excellent performance across all disciplines. 
Among the evaluated models, the fine-tuned \texttt{gemma-27b} achieved the highest F1-score at 93.5\%. %, followed by Gemma 9b and Mistral 22b.
Remarkably, the best-performing LLM fine-tuned on \textit{PEM-Rel-8K} achieved an average F1-score across the three disciplines that was only 1.2\% lower than that of models fine-tuned exclusively on a single discipline.
Furthermore, our analysis of cross-discipline adaptability indicates that LLMs trained on one discipline can generalise effectively to others. %, maintaining competitive performance across different domains.
These results demonstrate that \textit{PEM-Rel-8K} enables the development of robust models capable of generalising across multiple research areas, making them well-suited for constructing multidisciplinary research topic ontologies.

Future work will advance along four primary directions. 
First, we plan to extend our research to additional disciplines. We will begin with STEM fields and then progressively expand to the Social Sciences, Humanities, and Linguistics.
Second, we plan to incorporate cross-domain taxonomies, such as the Dewey Decimal Classification and the Library of Congress Subject Headings, to support cross-disciplinary alignment and to establish a standardised global framework for unifying specialised domain ontologies.
Third, we aim to develop an LLM-based system for ontology matching and evolution that integrates academic ontologies by identifying and extracting core hierarchical and synonymous relations. In this context, we also plan to develop an additional module to identify more nuanced relations, such as \textit{part-of}, \textit{instance-of}, and \textit{has-attribute}.
Finally, we plan to apply this system to integrate and extend a selection of prominent taxonomies, aiming to construct a comprehensive, multi-discipline ontology of research topics. We believe that such a solution has the potential to address current issues of coverage and fragmentation, thereby providing valuable support for repositories, digital libraries, academic search engines, and AI-powered tools.

\bibliographystyle{ACM-Reference-Format}
\bibliography{sample-base}

\end{document}

%% file: tables/llms.tex
\begin{table}[]
\centering
\caption{Overview of the 12 LLMs used in our experiments. The table includes the \textbf{Model} name, the alias adopted in this paper, the number of trainable \textbf{Parameters}, the context \textbf{Window} size, and the rank and scaling factor of the low-rank adaptation matrices used in LoRA (\textbf{r} and \textbf{alpha}).} 
\label{table:llms}
% \scriptsize 
\begin{tabular}{l|l|r|r|r|r}
\toprule
\textbf{Model}                       & \textbf{Alias}         & \textbf{Parameters} & \textbf{Window} & \textbf{r}   & \textbf{alpha} \\ \midrule
mistral-7b-instruct-v0.3    & mistral-7b     & 7.25B      & 32K    & 16  & 16    \\
Mistral-Nemo-Instruct-2407  & mistral-nemo-12b  & 12.2B      & 128K   & 16  & 16    \\
Mistral-Small-Instruct-2409 & mistral-22b & 22.2B      & 128K   & 16  & 16    \\
Llama-3.2-3B-Instruct       & llama-3b      & 3.21B      & 128K   & 256 & 128   \\
llama-2-7b-chat             & llama-chat-7b   & 7B         & 4K   & 256 & 128   \\
Meta-Llama-3.1-8B-Instruct  & llama-8b      & 8.03B      & 128K   & 256 & 128   \\
gemma-2b-it                 & gemma-2b      & 2.51B      & 8K   & 256 & 128   \\
gemma-2-9b-it               & gemma-9b      & 9.24B      & 8K   & 256 & 128   \\
gemma-2-27b-it              & gemma-27b     & 27.2B      & 8K   & 256 & 128   \\
Phi-3.5-mini-instruct       & phi-3         & 3.82B      & 128K   & 256 & 128   \\
phi-4                       & phi-4         & 14.7B      & 16K    & 256 & 128   \\
zephyr-sft                  & zephyr-7b        & 7.24B      & 8K     & 16  & 16   \\
\bottomrule
\end{tabular}
\end{table}

%% file: tables/transfer-table.tex
\begin{table}[!h]
% \caption{F1-score, Precision, and Recall for the best model across 24 experimental setups. \textbf{Experiment} indicates the method used: \textbf{STD} (Standard Prompting), \textbf{bCoT} (bidirectional Chain-of-Thought), \textbf{FT} (Fine-Tuning), and \textbf{TL} (Transfer Learning). \textbf{Training Set} and \textbf{Test Set} show the datasets used, and \textbf{Model} lists the top performer in each case.\label{table:transfer-table}}
\caption{F1-score, Precision, and Recall for the best-performing model in the 24 experimental configurations. The \textbf{Approach} column denotes the methodological strategy employed: \textbf{STD} refers to Standard Prompting, \textbf{bCoT} to bidirectional Chain-of-Thought, and \textbf{FT} to Fine-Tuning. The \textbf{Training Set} and \textbf{Test Set} columns specify the datasets used for model training and evaluation, respectively, while the \textbf{Model} column identifies the top-performing model in each configuration.\label{table:transfer-table}}
\centering
\begin{tabular}{@{}l|l|l|r|r|r|l@{}}
\toprule
\multicolumn{1}{c|}{\textbf{Approach}} & \multicolumn{1}{c|}{\textbf{Training Set}} & \multicolumn{1}{c|}{\textbf{Test Set}} & \multicolumn{1}{c|}{\textbf{F1-Score}} & \multicolumn{1}{c|}{\textbf{Precision}} & \multicolumn{1}{c|}{\textbf{Recall}} & \multicolumn{1}{c}{\textbf{Model}}  \\ \midrule
FT & IEEE-Rel-3K & IEEE-Rel-3K & {\ul \textbf{0.989}} & 0.989 & 0.989 & gemma-27b \\
FT & MeSH-Rel-4K & IEEE-Rel-3K & 0.945 & 0.947 & 0.945 & gemma-27b \\
FT & PhySH-Rel-875 & IEEE-Rel-3K & 0.947 & 0.947 & 0.947 & phi-4 \\
FT & PEM-Rel-8K & IEEE-Rel-3K & {\ul 0.973} & 0.974 & 0.973 & gemma-27b \\
STD & - & IEEE-Rel-3K & 0.716 & 0.830 & 0.747 & mistral-22b \\
bCoT & - & IEEE-Rel-3K & 0.769 & 0.809 & 0.787 & mistral-7b \\ \midrule
FT & IEEE-Rel-3K & MeSH-Rel-4K & 0.782 & 0.834 & 0.785 & mistral-22b \\
FT & MeSH-Rel-4K & MeSH-Rel-4K & {\ul \textbf{0.917}} & 0.918 & 0.917 & phi-4 \\
FT & PhySH-Rel-875 & MeSH-Rel-4K & 0.877 & 0.877 & 0.877 & gemma-27b \\
FT & PEM-Rel-8K & MeSH-Rel-4K & {\ul 0.908} & 0.911 & 0.907 & gemma-27b \\
STD & - & MeSH-Rel-4K & 0.669 & 0.766 & 0.694 & gemma-9b \\
bCoT & - & MeSH-Rel-4K & 0.716 & 0.775 & 0.725 & gemma-9b \\ \midrule
FT & IEEE-Rel-3K & PhySH-Rel-875 & 0.842 & 0.844 & 0.865 & phi-4 \\
FT & MeSH-Rel-4K & PhySH-Rel-875 & 0.864 & 0.870 & 0.865 & phi-4 \\
FT & PhySH-Rel-875 & PhySH-Rel-875 & {\ul \textbf{0.936}} & 0.946 & 0.930 & gemma-9b \\
FT & PEM-Rel-8K & PhySH-Rel-875 & {\ul 0.925} & 0.927 & 0.925 & gemma-27b \\
STD & - & PhySH-Rel-875 & 0.666 & 0.786 & 0.655 & mistral-22b \\
bCoT & - & PhySH-Rel-875 & 0.719 & 0.762 & 0.705 & mistral-7b \\ \midrule
FT & IEEE-Rel-3K & PEM-Rel-8K & 0.861 & 0.877 & 0.863 & mistral-22b \\
FT & MeSH-Rel-4K & PEM-Rel-8K & 0.919 & 0.920 & 0.919 & phi-4 \\
FT & PhySH-Rel-875 & PEM-Rel-8K & 0.906 & 0.907 & 0.906 & gemma-27b \\
FT & PEM-Rel-8K & PEM-Rel-8K & {\ul \textbf{0.935}} & 0.935 & 0.935 & gemma-27b \\
STD & - & PEM-Rel-8K & 0.677 & 0.779 & 0.703 & mistral-22b \\
bCoT & - & PEM-Rel-8K & 0.730 & 0.762 & 0.741 & mistral-7b \\ \bottomrule
\end{tabular}%
\end{table}

%% file: tables/pem-vs-pem.tex
\begin{table}[!ht]
\centering
% \caption{Precision, Recall, and F1-score for models fine-tuned on the \textit{PEM-Rel-8K} training set and evaluated on the \textit{PEM-Rel-8K} test set. BR refers to performance on \texttt{broader} relations, NA to \texttt{narrower}, OT to \texttt{other}, and SA to \texttt{same-as}. AVG indicates the average performance across the four categories. The best-performing scores for each relation are highlighted in {\ul\textbf{bold and underlined}}. Due to space constraints, we truncated the leading zero to all values.}
\caption{Precision, Recall, and F1-score for models fine-tuned on the \textit{PEM-Rel-8K} training set and evaluated on the \textit{PEM-Rel-8K} test set. BR refers to performance on \texttt{broader} relations, NA to \texttt{narrower}, OT to \texttt{other}, and SA to \texttt{same-as}. AVG indicates the average performance across the four categories. The best-performing scores for each relation are highlighted in {\ul\textbf{bold \& underlined}}. Due to space constraints, the leading zero has been omitted from all values.}
\label{table:pem-vs-pem}
\scriptsize
\begin{tabular}{l|ccccc|ccccc|ccccc}

\toprule
\multicolumn{1}{c|}{\multirow{2}{*}{\textbf{MODEL}}} & \multicolumn{5}{c|}{\textbf{F1-SCORE}} & \multicolumn{5}{c|}{\textbf{PRECISION}} & \multicolumn{5}{c}{\textbf{RECALL}} \\
\multicolumn{1}{c|}{} & AVG & BR & NR & OT & SA & AVG & BR & NR & OT & SA & AVG & BR & NR & OT & SA \\
\midrule
mistral-7b & .906 & .905 & .895 & .947 & .877 & .907 & .917 & .931 & .932 & .848 & .906 & .893 & .861 & .963 & .909 \\
mistral-nemo-12b & .907 & .902 & .886 & .949 & .890 & .907 & .883 & .912 & .96 & .875 & .907 & .922 & .861 & .939 & .906 \\
mistral-22b & .922 & .921 & .911 & .958 & .899 & .924 & {\ul \textbf{.963}} & .938 & .945 & .852 & .923 & .883 & .885 & .971 & {\ul \textbf{.953}} \\
llama-3b & .862 & .856 & .857 & .905 & .830 & .867 & .813 & .820 & .957 & .880 & .861 & .902 & .898 & .859 & .784 \\
llama-chat-7b & .887 & .891 & .865 & .932 & .861 & .888 & .915 & .877 & .911 & .847 & .888 & .868 & .854 & .954 & .875 \\
llama-8b & .910 & .909 & .895 & .946 & .889 & .910 & .900 & .918 & .936 & .886 & .910 & .919 & .873 & .956 & .891 \\
gemma-2b & .892 & .896 & .868 & .932 & .872 & .895 & .888 & {\ul \textbf{.943}} & .921 & .829 & .893 & .905 & .805 & .944 & .919 \\
gemma-9b & .926 & .919 & .917 & .958 & .909 & .926 & .919 & .932 & .943 & {\ul \textbf{.909}} & .926 & .919 & .902 & {\ul \textbf{.973}} & .909 \\
gemma-27b & {\ul \textbf{.935}} & {\ul \textbf{.935}} & {\ul \textbf{.922}} & {\ul \textbf{.966}} & {\ul \textbf{.916}} & {\ul \textbf{.935}} & .961 & .920 & .966 & .894 & {\ul \textbf{.935}} & .910 & {\ul \textbf{.924}} & .966 & .940 \\
phi-3 & .899 & .894 & .892 & .928 & .883 & .900 & .867 & .898 & .961 & .875 & .899 & .922 & .885 & .898 & .891 \\
phi-4 & .918 & .913 & .914 & .950 & .895 & .919 & .900 & .925 & {\ul \textbf{.970}} & .880 & .918 & {\ul \textbf{.927}} & .902 & .932 & .912 \\
zephyr-7b & .915 & .911 & .898 & .951 & .900 & .915 & .902 & .916 & .951 & .891 & .915 & .919 & .880 & .951 & .909 \\
AVG & .906 & .904 & .893 & .943 & .885 & .907 & .902 & .910 & .946 & .872 & .906 & .907 & .877 & .942 & .899 \\
\bottomrule
\end{tabular}%
% }
\end{table}